\documentclass{JAC2000}

\usepackage{graphicx}

\begin{document}
\title{HIGH POWER MODEL FABRICATION OF BIPERIODIC L-SUPPORT DISK-AND-WASHER STRUCTURE}
\author{H. Ao
\thanks{Present address: High Energy Accelerator Research Organization (KEK), 1-1 Oho, Tsukuba, Ibaraki, 305-0801, Japan},
 Y. Iwashita, T. Shirai and A. Noda, \\
Accelerator Laboratory, NSRF, ICR, Kyoto Univ., \\
M. Inoue, Research Reactor Institute, Kyoto Univ.,\\
T. Kawakita, K. Ohkubo and K. Nakanishi, Mitsubishi Heavy Industries, Ltd.}

\maketitle

\begin{abstract}
The high power test model of biperiodic L-support disk-and-washer was fabricated. Among some trouble in the fabrication, the main one was a vacuum leak in a brazing process. The repair test of the leak showed a good result; four units were recovered out of five leak units (recover rate 80\%). 
While an accelerating mode frequency was tuned at an operating frequency of 2857\,MHz by squeezing method, a coupling mode frequency of 2847MHz and the 3.4\% field flatness (peak to peak ratio) were achieved.
\end{abstract}
\section{Introduction}
A disk-and-washer (DAW) structure is developed as an advanced structure of a coupled cell cavity. The DAW structure has a high shunt impedance in the high $\beta$ region and good vacuum properties. A coupling constant of the DAW is much larger than that of a side-couple cavity, which brings easy frequency tuning and large tolerance in fabrication. The large coupling does not require a frequency tuning for each cell. Only the average frequency of entire cells must be controlled. Electric field distribution can be adjusted by slight movement of the washer position, because the distribution depends on the coupling constant balance between cells.

Higher mode acceleration complicates mode analysis for cavity design. The DAW requires a few supports which disturb an axial symmetry, so that the electromagnetic field of the DAW is more complicated than that of a side couple structure. Nevertheless, the feature of the DAW is attractive for a high $\beta$ region accelerator. This study of the DAW introduced the biperiodic L-support structure whose advantage was proposed by a calculation study.\cite{iwashita NIM} This study investigates its feasibility through the test model fabrication and the measurement. 

A coaxial bridge coupler connects two accelerator tubes of 1.2\,m length. The total length is about 2.8\,m. The operating frequency of this test model is 2857\,MHz which is the same as that of the disc-loaded linac in our facility. These specifications were designed so that a high-power test can be carried out with the existing beam line and RF sources in future. 

This paper describes the fabrication process of two acceleration tubes (No.1 and No.2) and the tuning operation.
\section{Fabrication}
\subsection{Fabrication process and structure} 
\label{Fabrication process and structure}
The DAW structure is fabricated by three brazing steps.(See fig.\ref{pic:3step})
\begin{figure}[htb]
\centering
\caption{Fabrication steps of the biperiodic L-support DAW}
\label{pic:3step}
\end{figure}

The wall loss on the metal surface requires water paths in the washer and support. Cooling water enters from one side of the supports and goes out from another side. Two rough processed half washers were brazed together (step1), and then it was machined to the final dimensions. 

The frequency and the field flatness of each unit were measured at the stage of step2 with aluminum units and terminating plates, which are used for the cold model test. These properties were optimized by fine corrections based on the measurement\cite{APAC1998}. A detail of the measurement and optimization are describes in \S \ref{Reguler_unit}.

The accelerator tube is installed into a water jacket made of SUS, which keeps the strength. 

\subsection{Brazing}
In this DAW structure, some brazed areas separate vacuum region from water. The reliability of the brazing is important for the DAW. The test model was fabricated in five times for the physical and technical R\&D (from 1st to 5th model generations). Following sections describe the problems and actions taken throughout this study.  
\subsubsection{Leak}
The fabricated model was inspected against vacuum leak in every brazing step. Although only the washers that passed the leak-test at step1 were used in step2, vacuum leaks arose in washer parts. Rates of the leak are summarized in Table \ref{Table:leaktest}. 
\begin{table}
\begin{center}
\begin{tabular}{ccc} 
\hline 
Model generation & Leak/Total & \% \\
\hline
1st & 0/6 & 0 \\
2nd & 5/8 & 63 \\
3rd & 6/12 & 50 \\
4th & 8/12 & 67 \\
5th & 0/12 & 0 \\
\hline
\end{tabular}
\caption{Leak-test results after brazing STEP2}
\label{Table:leaktest} 
\end{center}
\end{table}

Although the 1st and 5th models had no leak, the 2nd to 4th models exhibited bad yield (50 to 67\%).
 The reason is considered as follows. When fifty washers were brazed which were mainly used from the 2nd to 4th models, a vacuum leak arose on half of them. These washers were repaired by putting an additional brazing filler metal on the washer surface and brazing again. Although the leak was repaired at the time, the fine machining, as mentioned in \S \ref{Fabrication process and structure}, removed the surface and the brazed area became to thin. Even these washer parts passed the leak-test. Heating in step2, however, caused the vacuum leak again. 

The original reason for the washer repair is considered that they waited about one year after the machining till the step1 brazing. It caused oxidation on the surface of brazing area, which degraded the brazing quality.

The 1st and 5th model had no idle time, hence there is no vacuum leak.
\subsubsection{Leak repairing}
The five leak units were tested for repair, and the four of them were successful at the stage of step2. The way of repairing was as follows. After paste mixed with a brazing metal powder was injected into a cooling path, these units were heated up again. Because the paste metal must not fill up the cooling path, viscosity of the paste was adjusted by additional acetone, and compressed air was blown after inserting the paste. The repaired washer was set at lower side in a heat up process for the brazing metal not to flow out.

\begin{figure}[htb]
\centering
\caption{Repairing a vacuum leak by paste of brazing filler metal.}
\label{pic:leak}
\end{figure}

Although three units had no vacuum leak after an entire brazing (step3), one unit caused a small vacuum leak again. The leak rate was $5.0\times10^{-9}$[torr$l$/$s$]. Because the brazing metal for the repair has melting temperature at the middle between step2 and step3, the margin of the melting temperature was ensured. Endurance of the repaired area, however, seems not enough.

This structure requires three steps brazing, so that some improvements are required in the washer fabrication. It is necessary to reduce the waiting time from the fine machining to the brazing process.  As for a basic improvement, the inner structure have more strong brazed junction.

\section{Measurements of properties}
\subsection*{Regular units}\label{Reguler_unit}
\subsubsection{Frequency}
The accelerating and coupling mode frequencies were measured at step2. Figure \ref{pic:1unit} shows the measurement results as a scatter diagram.

\begin{figure}[htb]
\centering
\caption{Single unit measurements of frequencies.}
\label{pic:1unit}
\end{figure}

The frequency errors are small within the same model generations. It means that the fabrication process keeps good reproducibility.
It is important to examine the dependence of the frequency on the number of units for the design of a multi-cell-DAW. Many aluminum models were used to optimize dimensions\cite{JJAP2000}. Figure \ref{pic:12unit} shows the frequency convergence up to 12 units of the optimized test model.

\begin{figure}[htb]
\centering
\caption{Unit number dependence on the accelerating and coupling mode frequencies.}
\label{pic:12unit}
\end{figure}

The tolerance of the accelerating mode frequency is about 1\,MHz at this stage. Because the squeeze method can raise the accelerating mode frequency up to about 1.5\,MHz. The coupling mode frequency is about 10\,MHz lower than the operating frequency 2857\,MHz from this result. This 10\,MHz error is considered as tolerable, because the DAW has a large coupling constant, and thus the mode separation is about 30\,MHz.  

\subsubsection{Electric field distribution}
The electric field distribution depends on the displacement of the washer. This study chose the following way to keep the precision of the washer position.
\begin{enumerate}
\item A coordinate measuring machine measures the support (fixed in a flame) positions after step1 brazing. 
\item The support socket is machined on the washer surface so that the washer center fits on the beam line.
\item A carbon rod holds the centerline throughout the heat-up process in the step2 brazing. 
\end{enumerate}
The tolerance of the washer position is $\pm0.1$\,mm from a concentric center and a parallel position.

Two sources of assembling error are considered. One is the displacement in the heat-up process. The other is the miss handling after brazing. The carbon rod must be removed without washer displacement. This work was sometimes not easy. This is because a thermal expansion of copper might tighten the clearance between the beam-hole of the washer and the carbon rod. 

The holding scheme of carbon supports should be easy to remove after brazing. These assembling error in the beam-axis direction can be easily corrected.   It is not easy on the transverse direction. The assembling tolerance and the loose contact should be compromised. 

In the DAW, the flatness of the electric field distribution can be corrected without changing the frequencies. Figure \ref{pic:field} shows the example of the field correction. The field distribution was checked before whole brazing (step3) assembled temporarily. A coordinate measuring machine measured the washer positions, and then they were corrected by hand. The final field distribution is also shown in Fig. \ref{pic:field}. 
\begin{figure}[htb]
\centering
\caption{Correction of the electric field distribution and the final distributions of No.1 and No.2 accelerating tubes.}
\label{pic:field}
\end{figure}
\section{Conclusion}
The feasibility of the biperiodic L-support DAW was confirmed at S-band as the result of this study. The fabrication procedure was established; machining, brazing and assembling. We took care of following points. The unit number dependence on the frequency must be considered during the frequency optimization. This optimization would need some test model measurements. The brazing process is most important in the fabrication. The reliability of the washer should be kept for the three brazing steps. More R\&D's are needed for managements of brazing and improvements of the inner structure of the washer. A machining process is important for frequency control. The fine machining of NC turning center achieved the sufficient precision for the reproduction. The electric field distribution can be corrected by hand after assembling based on the measurement data, so that it is not necessary to keep the fine precision throughout the fabrication. This correction has no influence on the frequencies. The squeeze has the 1.5\,MHz tuning range. This tolerance would be enough to optimize the cavity dimensions considering the frequency convergence.

\end{document}